\documentstyle[epsfig,12pt]{article}

\newcommand{\bce}{\begin{center}} 
\newcommand{\ece}{\end{center}}
\newcommand{\beq}{\begin{equation}}
\newcommand{\eeq}{\end{equation}}
\newcommand{\bea}{\vspace{0.25cm}\begin{eqnarray}}
\newcommand{\eea}{\end{eqnarray}}

\newcommand{\br}{{\bf r}}

\newcommand{\ba}{\begin{array}}
\newcommand{\ea}{\end{array}}

\newcommand{\ket}[1]{| {#1} \rangle}

\newcommand{\doublespace}{
    \renewcommand{\baselinestretch}{1.6}\large\normalsize}

\def\lsim{\mathrel{\rlap{\lower4pt\hbox{\hskip1pt$\sim$}}
    \raise1pt\hbox{$<$}}}         
\def\gsim{\mathrel{\rlap{\lower4pt\hbox{\hskip1pt$\sim$}}
    \raise1pt\hbox{$>$}}}         

\def\beq{\begin{equation}}
\def\endeq{\end{equation}}
\def\arr{\begin{eqnarray}}
\def\endarr{\end{eqnarray}}
\makeindex

  \advance\oddsidemargin  by -1in
  \advance\evensidemargin by -1in
  \advance\topmargin      by -0.5in
\begin{document}

\vspace{2.0cm}

\begin{flushright}
\end{flushright}

\vspace{1.0cm}

\begin{center}
{\Large \bf 
Charm  of small $x$ neutrino DIS}.

\vspace{1.0cm}

{\large\bf R.~Fiore$^{1 \dagger}$ and V.R.~Zoller$^{2 \ddagger}$}

\vspace{1.0cm}

$^1${\it Dipartimento di Fisica,
Universit\`a     della Calabria\\
and\\
 Istituto Nazionale
di Fisica Nucleare, Gruppo collegato di Cosenza,\\
I-87036 Rende, Cosenza, Italy}\\
$^2${\it
ITEP, Moscow 117218, Russia\\}
\vspace{1.0cm}
{ \bf Abstract }\\
\end{center}
Due to the weak current non-conservation 
 the diffractive  excitation of  charm and strangeness
dominates the longitudinal structure function $F_L(x,Q^2)$ of
 neutrino DIS at small Bjorken  $x$. Based on the color dipole BFKL approach
 we report quantitative 
predictions for this effect in the kinematical range of the 
CCFR/NuTeV experiment. We comment on the relevance of our findings to 
experimental tests of PCAC.

\doublespace

\vskip 0.5cm \vfill $\begin{array}{ll}
^{\dagger}\mbox{{\it email address:}} & \mbox{fiore@cs.infn.it} \\
^{\ddagger}\mbox{{\it email address:}} & \mbox{zoller@itep.ru} \\
\end{array}$

\pagebreak


{\bf{1. Introduction.}} 
The neutrino deep inelastic scattering 
(DIS)
at small values of the  Bjorken variable  $x_{Bj}=Q^2/2m_N\nu$ 
provides  a useful tool for
 studies of 
fundamental
 properties of electro-weak (EW) interactions. In particular,  the analysis of 
neutrino-nucleon cross sections at vanishing four-momentum transfer 
squared, $Q^2$, can be used to test the hypothesis of  partial conservation 
of the  axial current (PCAC) in the kinematical region of high leptonic 
energy transfer, $\nu$, \cite{Jones,Fleming} 
(for theoretical introduction see \cite{KopMar}). 
 The partial conservation hypothesis \cite{Nambu}
connects  via Adler's theorem \cite{Adler}
the longitudinal structure function (LSF) at $Q^2\to 0$
 induced by the light-quark
 axial-vector current ($ud$-current)
with   the on-shell pion-nucleon total cross section,
\beq
F^{ud}_L(x,Q^2\to 0)= {f^2_\pi\over \pi}\sigma_\pi(\nu),
\label{eq:FLPCAC}
\eeq
where $f_{\pi}\simeq 130$ MeV is the pion decay
constant  (see
 \cite{FZAdler} for more discussion on the origin of Eq.(\ref{eq:FLPCAC})). 
To test the  Eq.(\ref{eq:FLPCAC})   the
structure function $F_2=F_L+F_T$ measured experimentally 
 is extrapolated down to $Q^2\to 0$ 
making use of  the fact
that the transverse structure function $F_T$ for $ud$-current 
vanishes  at $Q^2\to 0$. It is assumed that the contribution of the 
 charmed-strange (cs) current
can be neglected. However, in \cite{FZAdler} it has been pointed out that
 the  non-conservation of 
both   axial-vector and vector $cs$ currents 
leads to the abundant production  of  charm and strangeness at 
$Q^2\lsim m_c^2$
and for $\nu$ well  above the charm-strangeness mass  threshold.

In this communication  we  analyze the charged current (CC)  DIS
 in the color dipole (CD)  representation of the small-$x$ QCD \cite{NZ91,M} 
(for the review see \cite{HEBECKER})
  with particular emphasis on the role of charm and strangeness
 in the nucleon structure
probed by longitudinally polarized  electro-weak bosons. 
We quantify  the phenomenon of weak current non-conservation
in terms  of the 
light cone wave functions (LCWF) of $\ket{c\bar s}$ and $\ket{u\bar d}$  states
 in the Fock state
expansion for   the light cone EW boson. In Adler's regime of $Q^2\to 0$
the strong un-equality of masses of the charmed and strange quarks 
manifests its effects and the CD   analysis  reveals the
 ordering of dipole sizes
\bea  
m_c^{-2}< r^2 < m_s^{-2}
\label{eq:ORDER}
\eea 
typical of the DGLAP \cite{D,GL,AP} approximation. 
 The  multiplication of $\log$'s 
like 
\beq
\alpha_S\log(m_c^2/ m_s^2)\log(1/x)
\label{eq:LogLog} 
\eeq
to
higher orders 
of perturbative QCD
ensures the dominance of the charmed-strange component, $F_L^{cs}$, of the LSF
\beq
F_L=F_L^{ud}+F_L^{cs}
\label{eq:csud}
\eeq
 already at $x_{Bj}\lsim 0.01$, 
in the kinematical domain  covered by  the CCFR/NuTeV experiment \cite{FLCCFR}.
In presence  of charm and strangeness  the slope of $F_2$ at small $Q^2$
 changes dramatically
 thus complicating the  access to genuine PCAC component of $F_2$.
{\footnote{Different aspects of
 the CC inclusive and diffractive 
DIS have been discussed in 
\cite{Kolya92,KolyaNUDIS,BGNPZ2,MILTHOM}.}.  

{\bf 2. Dipole cross sections and  light-cone  density 
of $c\bar s$ states.}
When viewed in the laboratory frame the neutrino  DIS
at small $x_{Bj}$ derives from the absorption of the quark-antiquark,
${u\bar d}$
and ${c\bar s}$, Fock components of  the light-cone $W^+$-boson.
We  focus on the vacuum exchange dominated leading $\log(1/x)$ region
of $x\lsim 0.01$ where
the contribution of excitation of open 
charm/strangeness to the absorption cross section for longitudinal 
($\lambda=L$)
and transverse  ($\lambda=T$)
$W$-boson of virtuality $Q^2$,
is given by the color dipole
factorization formula 
\cite{ZKL,BBGG,NZ91}
\beq
\sigma_{\lambda}(x,Q^{2})
=\int dz d^{2}{\bf{r}}
|\Psi_{\lambda}(z,{\bf{r}})|^{2} 
\sigma(x,r)\,.
\label{eq:FACTOR}
\eeq
The interaction 
of the color dipole of size $\bf r$ with the target nucleon 
is described by the CD cross section $\sigma(x,r)$.
In the color dipole  approach the BFKL-$\log(1/x)$ evolution \cite{BFKL} 
of $\sigma(x,r)$
is described by the CD BFKL equation of Ref.\cite{NZZBFKL}. 
For qualitative estimates the  Double Leading Log Approximation (DLLA)
\cite{D,GL,AP} is suitable. Then,  for small dipoles  \cite{PHLA10}  
\bea
\sigma(x,r)\approx 
{\pi^2r^2\over N_c} \alpha_S(r^2)\int^{A/r^2}_{\mu_G^2} {dk^2\over k^2}
{\partial G(x,k^2)\over\partial\log{k^2}}\nonumber\\
\approx {\pi^2 r^2\over N_c}\alpha_S(r^2)
G(x,A/r^2),
\label{eq:SMALL}
\eea
where $G(x,k^2)=xg(x,k^2)$ is the 
gluon structure function  
and $A\simeq 10$ \cite{PHLA10}. We use the one-loop strong coupling
$\alpha_S(k^2)=4\pi/\beta_0\log(k^2/\Lambda^2)$ with $\Lambda=0.3$ GeV and
$\beta_0=11-2N_f/3$. In the numerical estimates we impose the infrared 
freezing, $\alpha_S(k^2)\leq \alpha_S^{fr}=0.8$.
For large dipoles, $r\gsim r_S$, $\sigma(x,r)$ saturates and 
the saturation scale, $r_S$, is as follows
\beq
r^2_S={A\over \mu_G^2},
\label{eq:Amug}
\eeq
where  $\mu_G=1/R_c$ is the inverse 
correlation radius of perturbative gluons.  
 From the lattice QCD  studies $R_c\simeq 0.2-0.3$ fm \cite{MEGGIO}.
Because $R_c$    
is small  compared to the typical range of strong interactions, the 
dipole
 cross section  evaluated with  the  decoupling of soft gluons gluons,
$k^2\lsim \mu_G^2$,
 would underestimate  the interaction strength for
 large color dipoles. In Ref.\cite{NPT,NSZZ} this missing strength
 was modeled by 
a non-perturbative, soft correction $\sigma_{npt}(r)$ to the 
dipole cross section $\sigma(r)=\sigma_{pt}(r)+\sigma_{npt}(r).$ 
Specific form of $\sigma_{npt}(r)$ was successfully tested against
 diffractive vector meson production data \cite{Vec}.

 Denoted by  
$|\Psi_{\lambda}(z,{\bf{r}})|^2$ in (\ref{eq:FACTOR}) is the light 
cone density of 
$c\bar s$ states with the $c$ quark 
carrying fraction $z$ of the $W^+$ light-cone momentum and 
$\bar s$ with momentum fraction $1-z$. 
In particular, $|\Psi_{L}|^2$ in Eq.(\ref{eq:FACTOR})  
is the incoherent sum of two terms,  the vector, $V_L$,  and the  axial-vector, $A_L$
 \cite{Kolya92,FZ1},
\beq
|\Psi_{L}(z,\br)|^2= |V_{L}(z,\br)|^2+ |A_{L}(z,\br)|^2
\label{eq:PSIL2}
\eeq
with \cite{Kolya92,FZ1}
\bea
|V_L(z,{\bf r})|^2
={{2\alpha_W N_c}\over (2\pi)^2 Q^2}\left\{\left[2Q^2z(1-z)
+(m_c-m_s)[(1-z)m_c-zm_s]\right]^2K_0^2(\varepsilon r)
\right.
\nonumber\\
\left.
 +(m_c-m_s)^2
\varepsilon^2 K^2_1(\varepsilon r)\right\}
\label{eq:RHOS1}\\
|A_L(z,{\bf r})|^2
={{2\alpha_W N_c}\over (2\pi)^2 Q^2}\left\{\left[2Q^2z(1-z)
+(m_c+m_s)[(1-z)m_c+zm_s]\right]^2 K_0^2(\varepsilon r)
\right.
\nonumber\\
\left.
 +(m_c+m_s)^2
\varepsilon^2 K^2_1(\varepsilon r)\right\},
\label{eq:RHOS2}
\eea
where $\alpha_W=g^2/4\pi$ and the weak charge $g$ is
related to the Fermi coupling constant $G_F$,
\beq
{G_F\over \sqrt{2}}={g^2\over m^2_{W}}.
\label{eq:GF}
\eeq 
Hereafter, $m_c$ and $m_s$ are the quark and antiquark masses\footnote{In 
this paper we deal with constituent quarks  in the spirit 
of  Weinberg \cite{Weinberg}.
The renormalization of the axial charge $g_A$ is neglected here  and
the ratio $g_A/g_V$ for constituent quarks is assumed to be the same as
for current quarks, $g_A=g_V=g$.} and 
$\varepsilon^2$ which controls the transverse size of $c\bar s$ and, 
with obvious substitutions,  
of $u\bar d$ dipoles  is as follows
\beq
\varepsilon^2=z(1-z)Q^2+(1-z)m_c^2+zm_s^2 
\label{eq:EPSILON2}
\eeq
The terms proportional to $K_0^2(\varepsilon r)$ and $K_1^2(\varepsilon r)$ 
describe  the 
quark-antiquark states with the angular momentum $L=0$ (S-wave) and $L=1$
(P-wave), respectively.
The weak current non-conservation shows up in  
 terms $\propto m_c^2/Q^2$ and $ m_s^2/Q^2$ which dominate both 
 the vector $|V_L|^2$ and axial-vector $|A_L|^2$
density of states at small $Q^2$. 
The P-wave component of $|\Psi_L|^2$  arises
only due  to the current non-conservation.

{\bf 3. Qualitative estimates. DLLA.}
The absorption cross sections  
for longitudinal EW bosons, $\sigma_{L}$, defined by the Eq.(\ref{eq:FACTOR})
can be converted into the structure function $F_L$,
\beq
F_L(x,Q^2)={Q^2\over 4\pi^2\alpha_{W}}\sigma_{L}(x,Q^{2}),
\label{eq:FL}
\eeq  
Let us start with $F^{cs}_L(x,Q^2)$  at  large $Q^2$. 
At $Q^2\gg m_c^2$ the P-wave component of 
 $|\Psi_L|^2$ proportional to $K_1^2(\varepsilon r)$  vanishes 
approximately  as $(m_c^2/Q^2)\log(Q^2/m_s^2)$ and 
 the structure function $F^{cs}_L$ is dominated by   
  the S-wave component  represented by 
the terms $\propto K_0^2(\varepsilon r)$.
The asymptotic behavior of the 
Bessel function, $K_{0,1}(x)\simeq \exp(-x)\sqrt{\pi/2x}$ makes the 
$\bf{r}$-integration in Eq.~(\ref{eq:FACTOR}) rapidly convergent 
at $\varepsilon r \gsim 1$. For  $Q^2\gg m_c^2$, 
the  product 
$\alpha_S(Q^2)G(x,Q^2)$ is  flat in $Q^2$. Then,
 integration over $\bf{r}$  yields a broad symmetric $z$-distribution
\bea
F^{cs}_L\sim Q^4\int_0^1 dz 
{z^2(1-z)^2\over \varepsilon^4}
\alpha_S(\varepsilon^2)G(x,A\varepsilon^2) \nonumber\\
\sim \alpha_S(\overline{\varepsilon^2})G(x,A\overline{\varepsilon^2}),
\label{eq:FLS}
\eea
where $\overline{\varepsilon^2}\sim Q^2/4$  corresponds to 
the ``non-partonic'' 
domain of $z\sim 1/2$. Similar to the LSF of the  
muon induced  DIS ($\mu$DIS)
\cite{D, Cooper, Dick}, 
the LSF  of neutrino DIS ($\nu$DIS) is 
dominated by $r^2\sim 1/Q^2$ and provides a direct  probe
of the gluon density $G(x,Q^2)$ \cite{KolyaNUDIS}. The S-wave component of 
$F_L^{cs}$ decreases with decreasing $Q^2$, as shown in 
Fig.~\ref{fig:fig1} by the solid line,
 but contrary to the $\mu$DIS it
does not vanish at $Q^2\to 0$ because of the current non-conservation
generated by the  mass terms in Eqs.(\ref{eq:RHOS1},\ref{eq:RHOS2}). 
Deviations from the symmetric $z$-distribution do not lead to any 
sizable effects and $F^{cs}_L$ 
in  Eq.(\ref{eq:FLS}) flattens   at 
 $Q^2\sim m_c^2$ (see Fig.~\ref{fig:fig1}).

At moderate $Q^2\lsim m_c^2$  the P-wave component of $F_L^{cs}$ 
(dashed line in Fig.~\ref{fig:fig1}) takes over.
The P-wave  density of $c\bar s$ states is more singular at $r\to 0$, 
$K_1(\varepsilon r)\sim 1/\varepsilon r$.
 Then,  integration  over $\bf{r}$ 
in Eq.~(\ref{eq:FACTOR}) leads to the  $z$-distribution
\begin{figure}[h]
\psfig{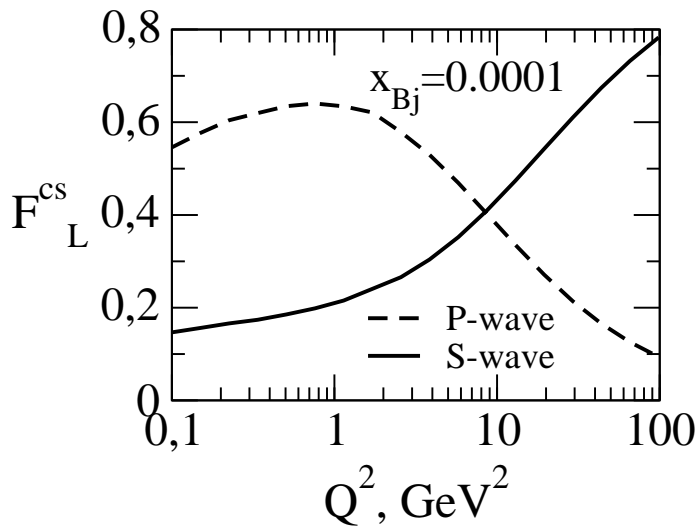}
\vspace{-0.5cm}
\caption{The  charm-strange component, $F^{cs}_L(x,Q^2)$,
 of the longitudinal neutrino-nucleon structure function.
Dashed curve corresponds to the P-wave
 contribution to $F_L^{cs}$,
solid curve represents the  S-wave component of $F_L^{cs}$.
The sum of two terms, $F_L^{cs}=P+S$,  is a slowly varying function of $Q^2$.} 
\label{fig:fig1}
\end{figure} 
\bea
{dF^{cs}_L\over dz}\sim 
{m_c^2\over Q^2z(1-z)+(1-z)m_c^2+zm_s^2}
\label{eq:FLP}
\eea
which develops  the parton model peaks at $z\to 0$ and $z\to 1$ 
at asymptotically large $Q^2$. At   $Q^2\lsim m_c^2$
the $z$-distribution becomes highly asymmetric, 
 only the peak at  $z\to 1$ survives, 
\bea
{dF^{cs}_L\over dz}\sim 
{1\over 1+\delta-z},
\label{eq:FLPNEW}
\eea 
where $\delta=m_s^2/(m_c^2+Q^2)$,
so that 
the charmed quark carries a fraction $z\sim 1-\delta$ of 
the  $ W^+$'s  light-cone momentum. With the Eq.(\ref{eq:FLPNEW}) 
the origin of $\log(m_c^2/ m_s^2)$
in (\ref{eq:LogLog}) becomes evident.

To clarify the issue of relevant dipole sizes
one can  integrate first over $z$
\bea
F_L^{cs} \sim {2N_c\over {(2\pi)^3}}m_c^2\int_0^1dz\int_0^{1/\varepsilon^2} 
{dr^2\over r^2}\sigma(x,r)\nonumber\\
\sim {2N_c\over {(2\pi)^3}}{m_c^2\over {m_c^2+Q^2}}
\int_{1/(m_c^2+Q^2)}^{1/m_s^2} 
{dr^2\over r^4}\sigma(x,r),
\label{eq:FLCSBORN}
\eea
where the factor $2$ is due to the additivity of $V$ and $A$ components of 
$F_L^{cs}$.
For numerical estimates we note that at $x\sim 0.01$
and moderate $Q^2$ the Born approximation (the two-gluon exchange) gives 
the results which are not unreasonable phenomenologically \cite{NZ91}.
For small dipoles the 2g-exchange yields
$\sigma(r)\sim \pi C_F\alpha_S^2 r^2\log(r_S^2/r^2)$ and the interpolating
function is
\beq
\sigma(r)\sim \pi C_F\alpha_S^2 r^2\log\left(1+{r_S^2\over r^2}\right).
\label{eq:2G}
\eeq
\begin{figure}[h]
\psfig{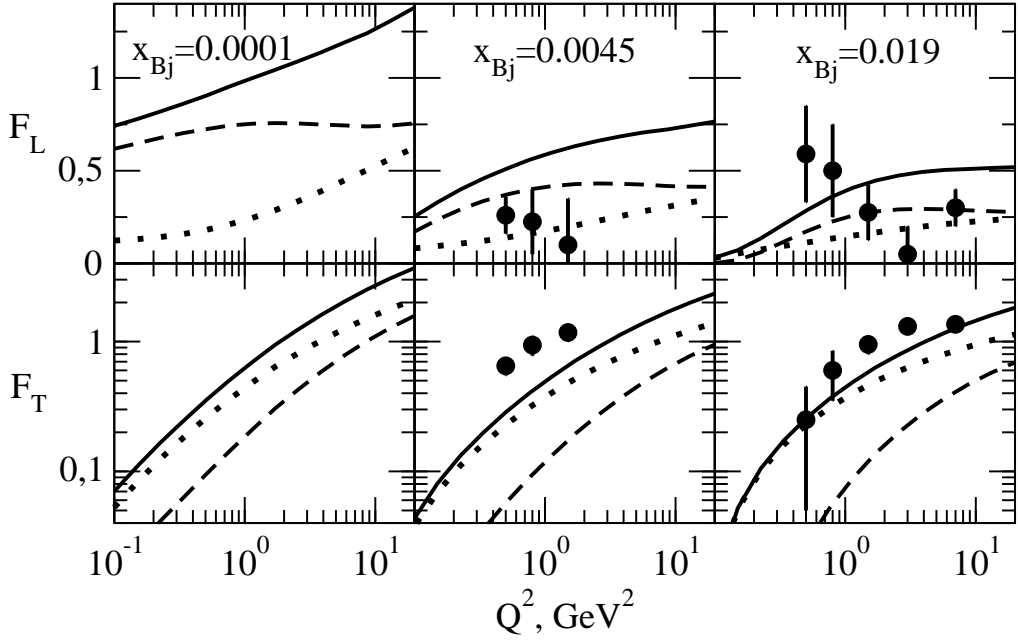}
\vspace{-0.5cm}
\caption{Data points are CCFR measurements of $F_L$ 
and $F_T=2xF_1$ \cite{FLCCFR}. 
Solid curves show the vacuum exchange contribution to  $F_L$ and $F_T$
in  $\nu Fe$ interactions.
Shown separately are the charm-strange (dashed curves) and light 
flavor (dotted curves) contributions to $F_L$ and $F_T$. 
Also shown are the predictions for $F_L$ and $F_T$
 at  $x_{Bj}=10^{-4}$.} 
\label{fig:fig2}
\end{figure}
Then, for the charmed-strange component of  $F_L$ 
 one gets
\beq
F_L^{cs}\sim {N_cC_F\over 4}{m_c^2\over {m_c^2+Q^2}}{1\over {2!}}L^2,
\label{eq:FL2G}
\eeq
where 
\beq
L={\alpha_S\over \pi}\log{\left(m_c^2+Q^2\over m^2_s\right)}
\label{eq:L}
\eeq
Here, $m_s^2$ introduces the infrared cutoff and stands, in fact, for 
  $max\{m_s^2,r_S^{-2}\}$ where  $r_S^{-2}$
comes from Eq.(\ref{eq:Amug}). 
In our numerical estimates the constituent strange quark mass equals to 
 $m_s=0.3$ GeV 
 and  is  close  to $r_S^{-1}$.

There is also a  contribution to 
$F_L^{cs}$  from the region $0<r^2<(m_c^2+Q^2)^{-1}$
\bea  
F_L^{cs}\sim {2N_c\over (2\pi)^3} 
m_c^2\int_0^1 dz\int^{1/(m_c^2+Q^2)}_0{dr^2\over r^2}
\sigma(x,r)\nonumber\\
\sim {N_cC_F\over 4}{m_c^2\over {m_c^2+Q^2}}
\left({\alpha_S\over \pi}\right)^2\log\left[r_S^2(m_c^2+Q^2)\right]
\label{eq:FLPEN1}
\eea
which is short of one $\log$, though. Notice the DLLA
ordering of sizes
\beq 
(m_c^2+Q^2)^{-1}<r^2<r^2_S,m_s^{-2}
\label{eq:ORDER1}
\eeq
announced in (\ref{eq:ORDER}) and elucidated by 
Eqs.(\ref{eq:FLCSBORN},\ref{eq:FLPEN1}). 

The rise of $F^{cs}_L(x,Q^2)$ towards small $x$ is generated by interactions
of the higher Fock states,  $c\bar{s}+gluons$, of the light-cone W-boson.
Making use of the technique developed in Ref.\cite{NZ91} one can estimate
the leading contribution to $F^{cs}_L$ associated with the Fock state 
 $c\bar{s}+one\,gluon$.  The result is 
\bea
\delta F_L^{cs}\sim {N_cC_FC_A\over 4}
{m_c^2\over {m_c^2+Q^2}}\log\left({x_0\over x}\right){1\over 3!}L^3
\label{eq:DELTAFL}
\eea
with $C_A\log(x_0/x)L$ as the DLLA expansion parameter.
The slope parameter $$\Delta={1\over 3}C_AL$$ is rather large,
$\Delta\simeq 0.3$ even at $Q^2=0$
 and we predict a rapid rise of $F^{cs}_L(x,Q^2)$ 
towards the region of small $x$.
\footnote{The DLLA resummation with the infrared cutoff $\mu_G$
results in 
$$
F_L^{cs}\sim {N_cC_F\over 4}{m_c^2\over 
{m_c^2+Q^2}}{L(m_c^2+Q^2)\eta^{-1}}I_2(2\sqrt{\xi}),
$$
where $I_2(z)\simeq \exp(z)/\sqrt{2\pi z}$ is the Bessel function,
 $\xi= \eta L(m_c^2+Q^2)$, 
$L(k^2)={4\over \beta_0}\log[{\alpha_S(\mu_G^2)/\alpha_S(k^2)}]$ 
and
$\eta=C_A\log\left({x_0/x}\right).$}

Our crude  estimate of the P-wave contribution
to $F^{cs}_L(x,0)$ given by Eqs.(\ref{eq:L},\ref{eq:FLPEN1},\ref{eq:DELTAFL}),
$F_L^{cs}+\delta F_L^{cs}\simeq 0.4$ at $x\simeq 10^{-4}$ and $x_0=0.03$, 
is  compatible  with the results of the CD 
BFKL analysis shown in Fig.~\ref{fig:fig1}. For comparison, Adler's theorem
gives for $F^{ud}_L(x,0)$ the value 
$f^2_\pi\sigma_\pi(\nu)/\pi\simeq 0.30-0.35$
and allows only a slow rise of $F^{ud}_L(x,0)$ with $\nu\propto x^{-1}$,
\beq
F^{ud}_L(x,0)\propto (1/x)^{\Delta_{soft}},
\label{eq:NU008} 
\eeq
where $\Delta_{soft}\simeq 0.08$ comes from the Regge parameterization of 
the total $\pi N$ cross section \cite{DonLan}.
Therefore, the   charmed-strange 
current dominates $F_L$ at   small-$x$. 

{\bf 4. Numerical results and discussion.} 
We evaluate $F_L$, $F_T$ and $F_2$, 
\beq
F_2(x,Q^2)=F_L(x,Q^2)+F_T(x,Q^2),
\label{eq:F2}
\eeq
for the $\nu Fe$ and $\nu Pb$ interactions 
  making use of the  
 approach to nuclear shadowing  developed in \cite{NSZZ}.
The $\log(1/x)$-evolution  
is described by the CD BFKL equation with boundary condition 
at $x_0=0.03$.
\begin{figure}[h]
\psfig{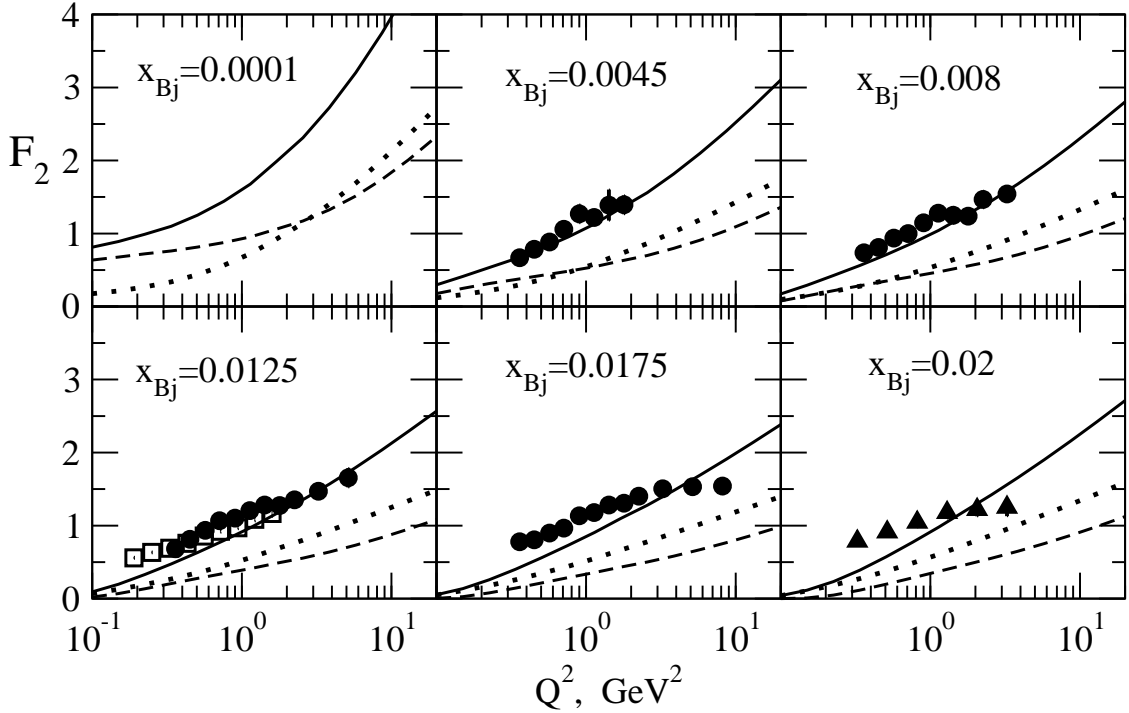}
\caption{
The nucleon structure function $F_2(x,Q^2)$ at smallest available  $x_{Bj}$
 as measured in   $\nu Fe$ CC DIS by the 
CCFR \cite{Fleming} (circles) and  
CDHSW Collaboration \cite{CDHSW} (squares, $x_{Bj}=0.015$). 
Triangles are  the CHORUS Collaboration  measurements  \cite{CHORUS}
of $F_2$  in 
 $\nu Pb$ CC DIS. 
 Solid curves show the vacuum  exchange 
contribution to $F_2(x,Q^2)$. 
Also shown are the charm-strange (dashed curves) and light 
flavor (dotted curves) components of $F_2$.} 
\label{fig:fig3}
\end{figure}
 In order to give a crude idea of
 finite energy effects at moderately small $x$ we stretch our estimates 
to $x>x_0=0.03$ multiplying the above CD cross sections by the purely 
phenomenological factor $(1-x)^5$ motivated by the familiar large-$x$ 
behavior of DGLAP parameterizations of the gluon structure function of the 
proton. Here $x$  makes sense of the gluon momentum fraction and equals to  
$x=x_{Bj}(1+M^2/Q^2)$ for $Q^2\lsim M^2$.
 For 
$Q^2\gsim M^2$,  $x=2x_{Bj}$ what corresponds to the
collinear DLLA. 
The mass scale $M$  differs for vector and axial-vector 
channels, $M=m_{\rho},m_{a_1}$ 
thus introducing non-universality of  $\sigma(x,r)$. For charmed-strange 
states we put $M^2=4$ GeV$^2$.

The CCFR Collaboration  measurements \cite{FLCCFR} of 
the structure function  $F_L$ and $F_T=2xF_1$ as a function of $Q^2$ 
for two smallest values of $x$ 
are shown in Fig.~\ref{fig:fig2}. 
From Fig. \ref{fig:fig2} it follows that we strongly overestimate 
$F_L$ and underestimate $F_T$ considerably.
However, the sum of two structure functions, $F_2=F_L+F_T$, shown in
Fig. \ref{fig:fig3} is in reasonable  agreement with data \cite{Fleming}.
 Also shown are the high statistics measurement of  $F_2$  from
 charged current $\nu Pb$ interactions
at smallest available $x_{Bj}=0.02$ by the CHORUS Collaboration \cite{CHORUS}
and $F_2$ as measured by the CDHSW 
collaboration \cite{CDHSW} in the $\nu Fe$ DIS
at $x_{Bj}=0.015$ (shown by squares in Fig. \ref{fig:fig3}).
The cs-component dominates the LSF $F_L$ 
 already at $x_{Bj}=0.0045$ and affects the slope of both  $F_L$ 
and $F_2$ at $Q^2\to 0$. Therefore,  the extrapolation 
of experimentally measured $F_2$ down to $Q^2\to 0$ 
 can hardly be used directly  to test PCAC. 
The cs-contribution to $F_2$ is 
quite considerable already at  
$x_{Bj}=0.0045$ and 
dominates $F_2$ at $x_{Bj}=10^{-4}$ for $Q^2\lsim m_c^2$ as 
shown in Fig. \ref{fig:fig3}.
The latter observation  is important for planned  tests of PCAC 
with high energy neutrino beams.

We underestimate $F_2$ at  moderately
small $x\gsim 0.01$ and small $Q^2$ (valence component is not included). 
Besides, in our analysis of small-$x$ phenomena  we rely upon the color
 dipole factorization (\ref{eq:FACTOR}) which is equally valid 
for small and large dipoles, in both  perturbative and non-perturbative 
domains. However,  two factors in  (\ref{eq:FACTOR}) have  different 
status. The CD cross section $\sigma(r)=\sigma_{pt}(r)+\sigma_{npt}(r)$
 is corrected for  the 
effects of soft physics while  the light-cone density of states 
$|\Psi_L(r)|^2$ is of purely perturbative nature.
Non-perturbative corrections to $|\Psi_L(r)|^2$ at small $Q^2$ may cause
observable effects.
In \cite{FZAdler} it has been found that the color dipole models successfully 
tested against the
 DIS  data from HERA  underestimate 
$F^{ud}_L(x,0)$ defined by the  Eq.(\ref{eq:FLPCAC}).
Particularly, our model with 
$m_u=m_d=0.15$ GeV reproduces  only half of the empirical  value  
$f^2_{\pi}\sigma_{\pi}/\pi$, not quite bad for the model evaluation
of the soft observable, although not satisfactory either. 
This may lead to the  deficit of  $F_2$ in the kinematical region of 
moderately  small $x$ dominated by the $ud$-current.

It is worth noticing that the nuclear absorption  is  weaker for 
charmed-strange  states. Therefore,  there is a specific nuclear  
enhancement of the 
charm production compared to the excitation of  light flavors.
The analysis of 
nuclear effects in the CC DIS 
will be published elsewhere.  

{\bf 5. Summary.}
We 
developed the  color dipole
description  of the phenomenon of charged  current non-conservation
in the neutrino DIS at small Bjorken $x$. 
 We quantified the effect
 in terms of the tree level
 light-cone wave functions and found  that 
 the charmed-strange component of the longitudinal 
structure function   is much larger than its light quark
component  already at $x\sim 0.01$. 
We  found also that
  the excitation of charm and strangeness dominates  the 
structure function 
$F_2(x,Q^2)$ at   $Q^2\lsim m_c^2$ and small enough $x$. 
A structure function analysis \cite{Fleming, CDHSW,CHORUS} of  neutrino
 DIS data  lends support  to our predictions.

\vspace{0.2cm} \noindent \underline{\bf Acknowledgments:}
The authors are indebted  to   N.N. Nikolaev and B.G. Zakharov
for useful comments. 
V.R.~Z. thanks  the Dipartimento di Fisica dell'Universit\`a
della Calabria and the Istituto Nazionale di Fisica
Nucleare - gruppo collegato di Cosenza for their warm
hospitality while a part of this work was done.
The work was supported in part by the Ministero Italiano
dell'Istruzione, dell'Universit\`a e della Ricerca and  by
 the RFBR grant 06-02-16905  and 07-02-00021.

\end{document}